\documentclass[a4paper,11pt]{article}
\usepackage{pos}
\newcommand{\be}{\begin{equation}}
\newcommand{\en}{\end{equation}}
\newcommand{\bea}{\begin{eqnarray}}
\newcommand{\ena}{\end{eqnarray}}
\newcommand{\bp}{\begin{pmatrix}}
\newcommand{\ep}{\end{pmatrix}}
\newcommand{\lbl}[1]{\label{eq:#1}}

\newcommand{\rf}[1]{(\ref{eq:#1})}

\newcommand{\fig}[1]{\ref{fig:#1}}

\newcommand{\braque}[1]{{\langle #1 \rangle}}

\newcommand{\bc}{\begin{center}}
\newcommand{\ec}{\end{center}}
\newcommand{\bt}{\begin{tabular}}
\newcommand{\et}{\end{tabular}}
\newcommand{\ba}{\begin{array}}
\newcommand{\ea}{\end{array}}

\newcommand{\im}{{\rm Im\,}}

\newcommand{\disc}{{\rm disc\,}}
\newcommand{\Kbar}{\bar{K}}
\newcommand{\mpi }{m_\pi}
\newcommand{\mk }{m_K}

\newcommand{\mpid}{m_\pi^2}

\newcommand{\mkd}{m_K^2}

\newcommand{\Kp}{{K^+}} 

\newcommand{\Kz}{{K^0}} 

\newcommand{\piz}{{\pi^0}} 
\newcommand{\pip}{{\pi^+}}
\newcommand{\pim}{{\pi^-}}
\newcommand{\nubar}{\bar{\nu}}
\newcommand{\vud}{V_{ud}}
\newcommand{\vus}{V^*_{us}}

\newcommand{\gapprox}{%
\mathrel{%
\setbox0=\hbox{$>$}\raise0.6ex\copy0\kern-\wd0\lower0.65ex\hbox{$\sim$}}}
\newcommand{\lapprox}{%
\mathrel{%
\setbox0=\hbox{$<$}\raise0.6ex\copy0\kern-\wd0\lower0.65ex\hbox{$\sim$}}}

\title{Dispersive description of the $K \to \pi \ell^+ \ell^-$
  radiative amplitudes} 

\author[a]{V\'eronique Bernard}
\author[a]{S\'ebastien Descotes-Genon}
\author[b]{Marc Knecht}
\author*[a]{Bachir Moussallam}

\affiliation[a]{Universit\'e Paris-Saclay, CNRS/IN2P3, Laboratoire Ir\`ene
  Joliot-Curie, 91405 Orsay, France }

\affiliation[b]{CNRS/Aix-Marseille Univ./Univ. de Toulon, Centre de Physique
  Th\'eorique (UMR 7332)\\
  CNRS-Luminy Case 907,13288 Marseille Cedex 9, France}

\emailAdd{veronique.bernard@ijclab.in2p3.fr}
\emailAdd{sebastien.descotes-genon@ijclab.in2p3.fr}
\emailAdd{marc.knecht@cpt.univ-mrs.fr}
\emailAdd{bachir.moussallam@ijclab.in2p3.fr}
  
\abstract{
We propose a description of the $K^+$, $K_S$ radiative decay form
factors $W_+$, $W_S$ based on general properties of analyticity and
unitarity.  Starting from the simple consideration of the asymptotic
behaviour of the two combinations $2W_+-W_S$ and $W_+ +W_S$ we derive
a dispersive representation involving only two parameters. Using the
rich experimental information on the $K\to3\pi$ amplitudes, extended
beyond the low energy region using the Khuri-Treiman formalism, we
show that the sign of the $W_+$ form factor is unambiguously
determined and its energy dependence can be well reproduced. We also
show that the yet unknown $\Delta{I}=1/2$ part of the $K_S \to
\pi^+\pi^-\pi^0$ can be determined from the value of
$W_+(0)+W_S(0)$. The possibility of fixing the sign of $W_S$ from
experiment is discussed.
}

\FullConference{13th International Conference on Kaon Physics (KAON2025)\\
8-12, September, 2025\\
Mainz, Germany\\}

\begin{document}
\maketitle

\section{Introduction}
Experimental studies of rare Kaon decays induced by flavour changing
neutral currents have made important progress in the last years. The
``golden mode'' $K^+\to\pip\nu\nubar$ was observed for the first time
by the E949 collaboration~\cite{E949:2008btt} and the branching
fraction has been measured with a precision of 35\% by
NA62~\cite{NA62:2021zjw}. The same collaboration has measured the
branching fraction and the energy distribution of the radiative decay
mode $K^+ \to \pi^+ \mu^+\mu^-$ with a statistics of 35K
events~\cite{NA62:2022qes}. They plan to collect as much as 100K
events in both the $\mu^+\mu^-$ and $e^+e^-$ modes by the end of the
present run. These two radiative modes are of particular interest for
probing sources of lepton flavour violation~\cite{Crivellin:2016vjc}.
Much less is known on the analogous radiative decays of the $K_S$, of
which only a few events have been
detected~\cite{NA481:2003cfm,NA481:2004nbc}. A knowledge of the
$K_S\to\pi^0\ell^+\ell^-$ amplitude is important in order to precisely
evaluate the amplitude of the (dominantly) CP-violating golden mode
$K_L\to\pi^0\ell^+\ell^-$ in the Standard Model (SM).

A theoretical framework is necessary in order to extract information
on the amplitudes from the experimental data. We will reconsider here
the approach based on the general non-perturbative properties of
unitarity/analyticity of the SM. Since the Kaon is light, only a few
contributions to the unitarity relation need to be considered. The
leading contribution is from the $\pip\pim$ state and is proportional {to the}
$J=1$ projection of the $K^\pm\to\pip\pim\pi^\pm$, $K_S\to\pip\pim\piz$
amplitudes (see eq.~\rf{discWpipi} below). The chiral calculation of the
$K\to\pi\ell^+\ell^-$ amplitudes at order $p^4$ performed in
ref.~\cite{Ecker:1987qi} correspond to linear approximations of the
$K\to3\pi$ amplitudes, while the ``beyond one-loop model'' (B1L) proposed in
ref.~\cite{DAmbrosio:1998gur} uses quadratic polynomial approximations of
$K\to3\pi$ fitted to the experimental data and thus includes
contributions of chiral order $p^6$.

\begin{figure}[ht]
  \centering
\includegraphics[width=0.50\linewidth]{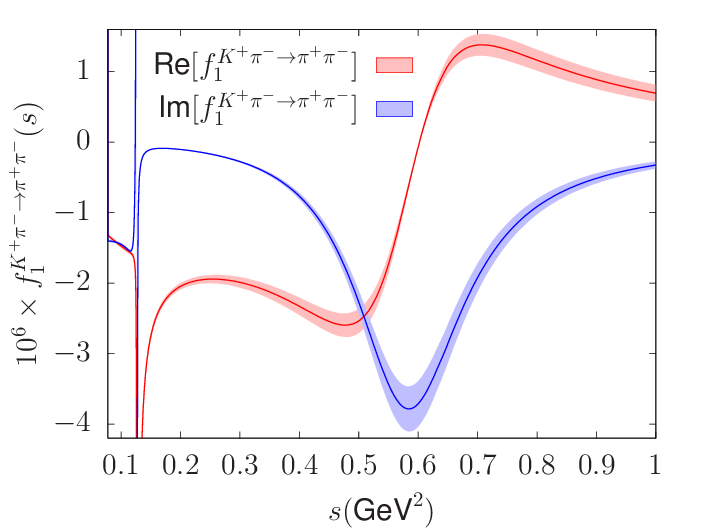}
\caption{\sl Real and imaginary parts of the $J=1$ projection of the
  $\Kp\to\pip\pim\pip$ amplitude from the KT formalism. The red and blue lines
  correspond to taking the central values of the KT subtraction parameters
  which are fitted to the experimental $K\to3\pi$ data. The error bands are
  generated by varying these parameters. One notices the presence of a
  singularity which occurs at the endpoint of the left hand (complex) cut at
  $s=(\mk-\mpi)^2$, which is integrable (e.g.~\cite{Kambor:1995yc} ).}  
\label{fig:f_1KPpi}
\end{figure}

These polynomial approximations are only valid inside the physical
K decay region. Instead, we propose here to use representations of the
$K\to3\pi$ amplitudes which derive from solutions of the integral
equations of Khuri and Treiman~\cite{Khuri:1960zz} (KT). These are valid
both in the decay and in the scattering regions in the energy range in
which $\pi\pi$ scattering is elastic. 
Fig.~\fig{f_1KPpi} illustrates the $J=1$ projection of the
$\Kp\to\pip\pim\pip$ amplitude evaluated with the KT solutions from
ref.~\cite{Bernard:2024ioq} showing the presence of the $\rho(770)$
resonance. In addition to $\pip\pim$ we will also include
the $K^0\piz$, $\Kp\pim$ channels. This proves necessary in order to obtain
a representation consistent with the cancellation of the UV divergences
and also allows to include the $K^*(892)$ resonance.
Examining the asymptotic behaviours, we will
argue that a minimal dispersive representation should hold involving
only two undetermined parameters (instead of four in the B1L model).

\section{Properties of the Kaon radiative amplitudes}
As is well known, radiative FCNC amplitudes in the SM are generated
from one-loop $\gamma$-penguin, $Z$-penguin and $W$-box diagrams
which, at low energy, can be represented by: a) a local part
represented by the two operators $Q_{7V}(x)$, $Q_{7A}(x)$
(e.g.~\cite{Buras:1994qa} and references therein)
\be
{\cal L}_{SD}= - \frac{G_F}{\sqrt{2}} \vud\vus \left[C_{7V}(\nu) Q_{7V}(x)
+ C_{7A} Q_{7A}(x)\right]
\en
with
\be
Q_{7V}= [\bar{s}\gamma_\mu(1-\gamma_5){d}]\, \bar{\ell}\gamma^\mu\ell  ,\quad
Q_{7A}=[\bar{s}\gamma_\mu(1-\gamma_5){d}]\, \bar{\ell}\gamma^\mu\gamma^5\ell 
\en
and b) a non-local part, which can be viewed as proceeding via
emission of a virtual photon, $K\to\pi\gamma^*$. This part can be
expressed as a matrix element involving the electromagnetic current 
$j_\mu^{em}=\frac{2}{3}\bar{u}\gamma_\mu{u}-\frac{1}{3}(
\bar{d}\gamma_\mu{d}+\bar{s}\gamma_\mu{s})$ and the $\Delta{S}=1$ effective
Lagrangian
\be\lbl{W+WSnu}
\ba{l}
i\int d^4 x 
\braque{\pi^+(p_2)\vert T \{ j_\mu^{em}(0) 
 \mathcal{L}_{\Delta{S}=1}(x) \} \vert K^+(p_1) }\equiv
\dfrac{2W_+(s;\nu)}{16\pi^2\mkd} Q_\mu ,\\[0.3cm]
i\int d^4 x 
\braque{\pi^0(p_2)\vert T \{ j_\mu^{em}(0) 
 \mathcal{L}_{\Delta{S}=1}(x) \} \vert K^0(p_1) }\equiv
\dfrac{\sqrt2W_S(s;\nu)}{16\pi^2\mkd} Q_\mu ,\\[0.3cm]
\ea
\en
where $Q_\mu=\frac{s}{2}(p_1+p_2)_\mu-\frac{\Delta_{K\pi}}{2}(p_1-p_2)_\mu$,
$s=(p_1-p_2)^2$ is the photon virtuality, $\Delta_{K\pi}=\mkd-\mpid$ and
$\mathcal{L}_{\Delta{S}=1}$ is expressed in terms of the six independent
$\Delta{S}=1$ four-quark operators~\cite{Buras:1994qa},
\be
\mathcal{L}_{\Delta{S}=1}(x)=-\frac{G_F}{\sqrt2}\vud\vus\sum_1^6 C_iO_i(x)\ .
\en
The amplitudes in eq.~\rf{W+WSnu} depend on the renormalisation scale
$\nu$, which reflects the short-distance singularity of the
$T$-product~\cite{DAmbrosio:1998gur,DAmbrosio:2018ytt}. The physical
form factors are obtained by adding the local and non-local
contributions,
\be
W_{+,S}(s)=W_{+,S}(s;\nu)\pm4\pi\frac{G_F\mkd \vud\vus}{\sqrt2\alpha}C_{7V}(\nu)
f_+^{K\pi}(s)\ .
\en
In this expression, isospin symmetry was used to express the matrix elements
of $Q_{7V}$ in terms of the semi-leptonic $K\pi$ vector form factor
$f_+^{K\pi}(s)$. 

Let us now consider the asymptotic behaviour of the form factors which is
important, obviously, in order to ensure convergence of the
dispersive integral representations. For this purpose, we can perform
the short distance expansion of the $T$-product
$T[j^{em}_\mu(0)\mathcal{L}_{\Delta{S}=1}(x)]$ which appears in
eq.~\rf{W+WSnu}. The leading term in this expansion, to lowest order
in $\alpha_s$, reads~\cite{DAmbrosio:2018ytt}
\be\lbl{OPEleading}
\lim_{q\to\infty}\int d^4x \, e^{iqx} T\{j_\mu^{em}(x)
  \mathcal{L}_{\Delta{S}=1}(0)\}
= (q_\mu q_\nu-g_{\mu\nu} q^2)(A+ B\ln\frac{-q^2}{\nu^2})
[\bar{s}\gamma_\nu(1-\gamma^5){d}]   
\en
involving the dimension 3 operator
$\bar{s}\gamma_\nu(1-\gamma^5){d}$. This operator carries isospin
$I=1/2$: It is useful, then, to consider the two combinations of the
form factors $W_{+,S}$ which correspond to $K\pi$ states having
$I=1/2$ and $I=3/2$,
\be
W^{[1/2]}\equiv 2W_+ -W_S,\quad  W^{[3/2]}\equiv W_+ + W_S\ .
\en
The leading term~\rf{OPEleading} in the OPE only affects the
combination $W^{[1/2]}$ and one can deduce that the ratio
$W^{[1/2]}/f_+^{K\pi}$ should satisfy a dispersive representation with
one subtraction constant. The large $s$ behaviour of $W^{[3/2]}$ is
affected by sub-leading terms in the OPE, which have dimension at
least 6. It is expected to satisfy a convergent representation with no
subtraction.

\section{Unitarity }
\noindent\underline{{\bf a) $\pip\pim$ }}: The contribution from $\pip\pim$ to the unitarity
relation of the form factors reads
\be\lbl{discWpipi}
\ba{l}
\dfrac{\disc[W_+(s)]_{\pi\pi}}{16\pi^2\mkd}=\dfrac{\theta(s-4\mpid)}{16\pi}
\left(1-\dfrac{4\mpid}{s}\right)^{3/2} [F_V^{\pi\pi}(s)]^* \times
f_1^{\Kp\pim\to \pip\pim} (s),
\\[0.4cm]
\dfrac{\disc[W_S(s)]_{\pi\pi}}{16\pi^2\mkd}=\dfrac{\theta(s-4\mpid)}{16\pi}
\left(1-\dfrac{4\mpid}{s}\right)^{3/2} [F_V^{\pi\pi}(s)]^*\times
f_1^{K_S\piz\to \pip\pim} (s) ,
\ea\en
where $F^{\pi\pi}_V(s)$ is the electromagnetic form factor of the pion,
\be
\braque{\pip(p_2)\vert j_\mu^{em}(0)\vert \pip(p_1)}=(p_1+p_2)_\mu
F_V^{\pi\pi}((p_1-p_2)^2) ,
\en
and
$f_1^{K\pi\to\pi\pi}$ are the $J=1$ partial-wave projections of the
$K\pi\to\pi\pi$ amplitudes.  The $K^+$ amplitude is entirely
determined in the physical decay region and will be extrapolated up to
the resonance region as shown in fig.~\fig{f_1KPpi}. The $K_S$
amplitude, in contrast, is only partly known. In this case, only the
component induced by $\Delta{I}=3/2$ part of the Lagrangian
$\mathcal{L}_{\Delta{S}=1}$ has been measured.  The remaining
component, induced by the $\Delta{I}=1/2$ part, is kinematically
suppressed in the physical decay region~\cite{Zemach:1963bc} and has
not yet been detected.  In the KT formalism, this $\Delta{I}=1/2$
amplitude is proportional to a single subtraction parameter
$\tilde{\mu}_1$ such that one can write
\be
f^{K_S\piz\to\pip\pim}_1= f^{K_S\piz\to\pip\pim}_{1,\Delta{I}=3/2}
+\tilde{\mu}_1 \bar{f}^{K_S\piz\to\pip\pim}_{1,\Delta{I}=1/2}
\en
where $\bar{f}^{K_S\piz\to\pip\pim}_{1,\Delta{I}=1/2}$ is known from
the KT solutions. This amplitude is suppressed at low energy but not
in the resonance region (a very
similar remark was made previously in ref.~\cite{Akdag:2023pwx}
studying the $\eta\to\piz\gamma^*$ CP-violating amplitude).
Using the integral representations, this will allow us to derive
a linear relation between the 
parameter $\tilde{\mu}_1$ and the value at $s=0$ of the form factors
$W_+(0)+W_S(0)$.

\vspace{0.4cm}
\noindent\underline{{\bf b) $\Kp\pim$, $\Kz\piz$ }}:
The $K\pi$ contributions to unitarity written for the isospin form
factors $W^{[1/2]}$, $W^{[3/2]}$, have a simple expression in the
elastic $K\pi$ scattering region, in terms of the corresponding
elastic $J=1$ phase-shifts $\delta_1^{1/2}$, $\delta_1^{3/2}$,
\be\lbl{unitpiK}
\disc[W^{[I]}(s)]_{{K\pi},\,{\rm elastic}}
=\exp(-i\delta_1^I(s))\sin(\delta_1^I(s)) 
  \,W^{[I]}(s)\ ,\quad I=1/2,\ 3/2
\en
These $K\pi$ contributions introduce a coupling between the $W_+$ and the
$W_S$ form factors.

\section{Dispersive integral representations}
We assume that the form factors $W_+$ and $W_S$ satisfy the usual
analyticity properties, i.e. that they are analytic functions of the
variable $s$ with cuts on the real axis associated with the unitarity
relations. We can separate the $\pi\pi$ cut from the $K{\pi}$ one by
appending an infinitesimal imaginary part to the kaon mass and write
Cauchy integral representations with a contour which turns around the
two cuts. The expression of the discontinuity across the $K\pi$
cut~\rf{unitpiK} implies that these have the form of
Muskhelishvili-Omn\`es integral equations which are solved in terms of
the Omn\`es functions $\Omega_1^I(s)$
\be\lbl{MOint}
\Omega^I_{1}(s)=\exp\left[ \frac{s}{\pi}\int_{m_+^2}^\infty ds'
  \dfrac{\delta_1^I(s')}{s'(s'-s) }
  \right]\ 
\en
with $m_+=\mpi+\mk$.
The phase in the integral~\rf{MOint} is given by the $K\pi\to{K}\pi$
$P$-wave phase-shift in the elastic scattering region $s'\le
\Lambda^2$, with $\Lambda\simeq1.2$ GeV, and can be chosen arbitrarily
above. In the case of $I=3/2$, the phase is very small and it is taken
to go to zero smoothly when $s' > \Lambda^2$. Dividing the form factor
$W^{[3/2]}$ by $\Omega_1^{3/2}$ removes the $K\pi$ cut and we arrive
at the simple dispersive representation in terms of the $\pi\pi$
discontinuities, 
\be\lbl{W32}
W^{[3/2]}(s)=
\Omega^{3/2}_{1}(s)\times\frac{1}{\pi}\int_{4\mpid}^{\Lambda^2} ds'
\frac{\disc[W_+(s')+W_S(s')]_{\pi\pi}}{(s'-s) \Omega^{3/2}_{1}(s')}   ,
\en
We made the assumption here that the contribution form the higher energy
region $s'> \Lambda^2$ can be neglected compared to that from the $\rho(770)$
resonance region. 

In the case of the $I=1/2$ form factor, one notes that the form factor
$f_+^{K\pi}(s)$ can be used as an Omn\`es function since it satisfies
the same phase representation~\rf{MOint}. We first introduce the
following once-subtracted dispersive integral over the $\pi\pi$
discontinuity
\be\lbl{W12_pipidef}
W^{[1/2]}_{\pi\pi}(s)\equiv \frac{s}{\pi}\int_{4\mpid}^{{\Lambda^2}} ds'
\frac{\disc[2W_+(s')-W_S(s')]_{\pi\pi}}{s'(s'-s) }
\en 
which carries the information on the $\rho(770)$ resonance. Next, we consider
the function $(W^{[1/2]}(s)- W^{[1/2]}_{\pi\pi}(s))/f_+^{K\pi}$ which can
be written in terms of the $K\pi$ discontinuity. Using one subtraction, we
arrive at the following representation for  $W^{[1/2]}(s)$ 
\be\lbl{W12nu}
W^{[1/2]}(s)=  W^{[1/2]}_{\pi\pi}(s)
+f_+^{K\pi}(s)\Big[G_F\mkd\dfrac{2a_+-a_S}{f_+^{K\pi}(0)}
-\frac{s}{\pi}\int_{m_+^2}^{\Lambda^2} ds'
\frac{W^{[1/2]}_{\pi\pi}(s')\im[1/f_+^{K\pi}(s')]}{s'(s'-s)}\Big]\  .
\en
where $a_+, a_S$, following the notation of the B1L
model~\cite{DAmbrosio:1998gur}, are proportional to the values at $s=0$
\be
W_+(0)\equiv G_F\mkd a_+,\quad W_S(0)\equiv G_F\mkd a_S
\en

\section{Real and imaginary parts of the radiative amplitudes}  
Adopting a sign convention for the $K\to3\pi$ amplitudes at low energy
fixes the signs of the $\pi\pi$ discontinuities
$\disc[W_{+,S}]_{\pi\pi}$ (see eq.~\rf{discWpipi}), and this opens the
possibility to determine the signs of the real and imaginary parts of
the amplitudes from the experimental data. Indeed, fits to the
experimental $K^+$ energy distribution based on the B1L model tends to
favour $a_+<0$ (this is reviewed in ref.~\cite{DAmbrosio:2018ytt}). In
order to probe the results from the dispersive approach against
experiment, we will simply use central values for $|a_+|$ and $|a_S|$
taken from refs.~\cite{NA62:2022qes,NA481:2003cfm},
\be\lbl{moda+aS} 
|a_+|= 0.575\pm0.012,\quad
|a_S|= 1.06\pm0.27\ .
\en

\begin{figure}[hbt]
\centering  
\includegraphics[width=0.49\linewidth]{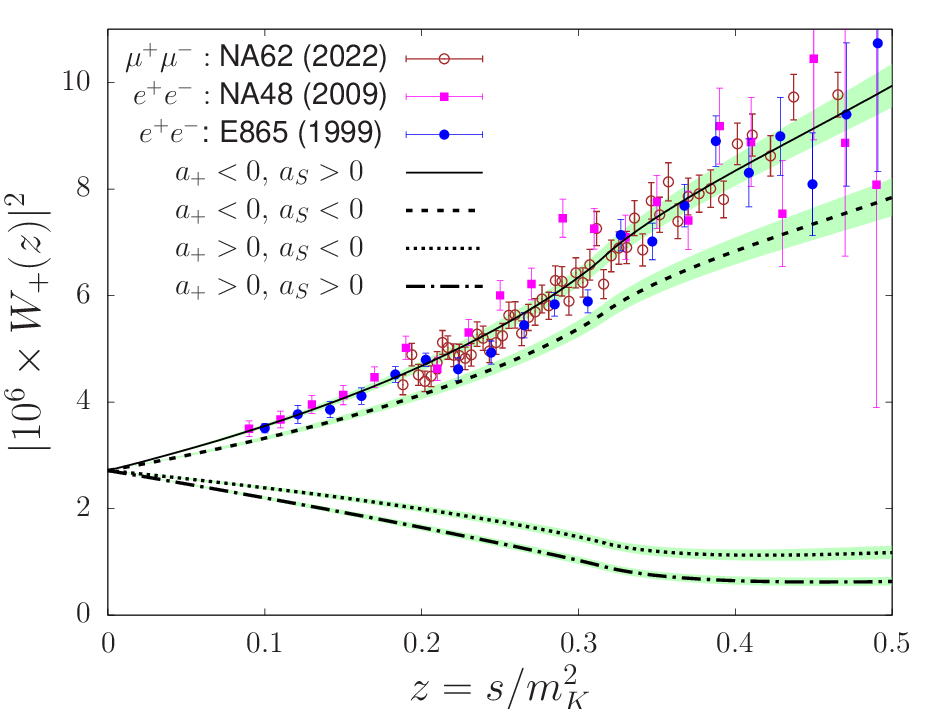}%
\includegraphics[width=0.49\linewidth]{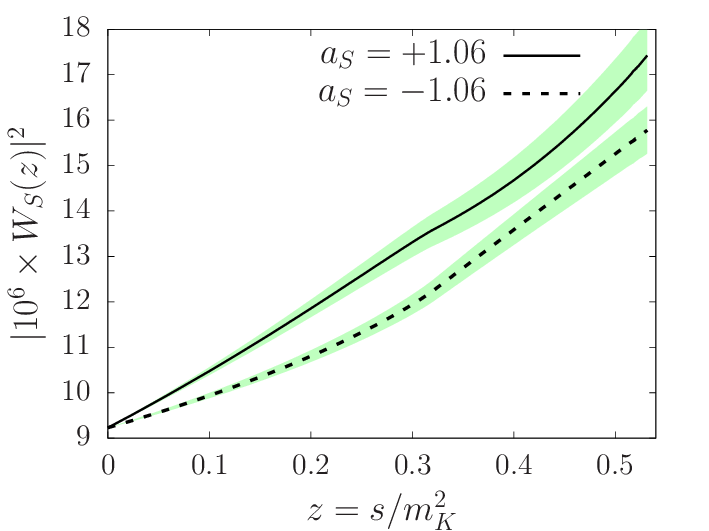}
\caption{\sl Left: Results for $|W_+|^2$ from the dispersive representations
  corresponding to different choices of signs for $a_{+,S}$: $a_+=\mp 0.575$,
  $a_S=\pm1.06$, compared to the experimental data from
  refs.~\cite{Appel:1999yq,Batley:2009aa,NA62:2022qes}. Right:
  Results for $|W_S|^2$ with $a_S>0$ and $a_S<0$.}
\label{fig:W2_plus}
\end{figure}

Evaluation of the dispersive integrals requires inputs for the form
factors $F_V^{\pi\pi}$ and $f_+^{K\pi}$. We have used phase
dispersive representations constrained by experimental data on
$e^+e^-\to\pip\pim$ scattering and $\tau\to{K}\pi\nu$ as well as
$K_{l3}$ decays and experimental results for the $\pi\pi$ and $K\pi$
phase shifts (see~\cite{Bernard:2025xyn} for details). The $K\to3\pi$
amplitudes are expressed as linear combinations of the set of
independent KT solutions computed in ref.~\cite{Bernard:2024ioq} with
coefficients determined from the experimental data. The one
undetermined coefficient $\tilde{\mu}_1$ can be evaluated, as
mentioned above, in terms of $a_++a_S$ using eq.~\rf{W32} at $s=0$,
which gives 
\be\lbl{Wsum2s=0}
\ba{ll}
10^8\,G_F\mkd\,(a_+ + a_S)= & -952.4\pm30.7+i(2.17\pm0.82) \\
 & +10^8\,\tilde{\mu}_1\,(0.73\pm0.04+i(1.12\pm1.39)\cdot10^{-3})\ ,
\ea\en
where the central values correspond to a cutoff $\Lambda=1.2$
GeV. Using this, one can compute the two form factors $W_+$, $W_S$
from the integral representations~\rf{W32}~\rf{W12nu},
given the values of $a_+$, $a_S$ as two inputs. The four curves on
fig.~\fig{W2_plus} (left) show the results for $|W_+(s)|^2$
corresponding to the four different sign choices compatible with
eq.~\rf{moda+aS}. The two curves which correspond to $a_+ >0$ have an
energy dependence in disagreement with experiment, such that a
positive sign for $a_+$ can be clearly excluded.
The energy dependence of $|W_+|^2$ has also a visible sensitivity to
the value of $a_S$. Comparison with the experimental data favours
$a_S>0$ and, in this case, a reasonable agreement with the data is
achieved with the values given in eq.~\rf{moda+aS}.

In order to test the model more completely, measurements of the energy
dependence of $|W_S|^2$ would be necessary. Such measurements could be
performed at LHCb in the $\mu^+\mu^-$ mode in the near
future. Figure~\fig{W2_plus} (right) shows the predicted energy
variation of $|W_S|^2$ from the dispersive model in the two cases of
$a_S=+1.06$ and $a_S=-1.06$ (with $a_+$ fixed to $a_+=-0.575$)
displaying a small but visible difference.
A combined fit to both $|W_+|^2$ and $|W_S|^2$
should enable one to fix the sign of $a_S$.

\section{Isoscalar resonance contributions}
The $\pi\pi$ and $K\pi$ unitarity contributions discussed above allow
one to take into account the effects of the light resonances
$\rho(770)$ and $K^*(892)$ in the form factors. Further 
resonances could play a role, in particular the light $I=0$ resonances
$\omega(783)$ and $\phi(1020)$. While these are often found to be
less important than the $\rho$ and $K^*$
(e.g.~\cite{Bergstrom:1990vj}), it turns out to be difficult to
evaluate their role precisely. In the unitarity approach, they can be
associated with $3\pi$ and $K\Kbar$ contributions respectively, but
these cannot be related to observable Kaon decays. An estimate was
made in ref.~\cite{Bernard:2025xyn} relying on the resonance weak
chiral Lagrangian proposed in ref.~\cite{Ecker:1992de}, which shows
that one cannot rule out that the $I=0$ resonances could modify the
slopes of the form factors by as much as 30\%, but this result is
sensitive to unknown terms of higher order in the resonance
Lagrangian.

From a more model independent point of view, one can rely 
on the $\Delta{I}=1/2$ rule which is well established for K decays.
According to this , the $I=0$
resonances should eventually affect the amplitude $W^{[1/2]}$ but not
$W^{[3/2]}$. If necessary, then, one could introduce one additional
subtraction parameter in the $W^{[1/2]}$ dispersive representation to
account for the isoscalar resonances at low energies.

\section{Summary}
We have developed a unitarity/analyticity approach to the form factors
$W_S$, $W_+$ describing the $K\to\pi \ell^+\ell^-$ decays which
goes  beyond previous work, in particular, through the use of $K\to3\pi$
amplitudes in the unitarity relations which can be extrapolated up to
the resonance region. We obtain a description which accounts for the
coupling between the two form factors induced by
$\Kp\pim\leftrightarrow\Kz\piz$ rescattering. In principle, this
allows one to derive information on $W_S$ based on $W_+$ which is easier to
measure. We argued that a minimal dispersive representation could
hold which involves only two undetermined parameters, $a_+$ and $a_S$. We
showed that it can describe rather well the experimental data on $|W_+|^2$ and
gave  predictions for the energy dependence of $|W_S|^2$ which
could be probed at LHCb.


\providecommand{\href}[2]{#2}\begingroup\raggedright\endgroup

\end{document}